\input harvmac
\input epsf

%%%%%%%%%%%%%%%%%%%%%%%%%%%%%%%%%%%%%%%%%%%%%%%%%%%%%%%%%%%%%%%%%%%%%%
%%%%%%%%%%%%%%%%%%%%%%%        references         
%%%%%%%%%%%%%%%%%%%%%%
%%%%%%%%%%%%%%%%%%%%%%%%%%%%%%%%%%%%%%%%%%%%%%%%%%%%%%%%%%%%%%%%%%%%%%

\lref\rBFSS{T. Banks, W. Fischler, S.H. Shenker and L. Susskind,
{\it M Theory as a Matrix Model: A Conjecture}, Phys. Rev. D55, 
(1997) 112.}
\lref\rTaylor{W. Taylor, {\it D-brane field theory on compact 
spaces,} 
Phys. Lett. B394 (1997) 283, hep-th 9611042.}
\lref\rTwn{P.K.Townsend, {\it Membrane tension and manifest $IIB$ $S$ 
duality,} hep-th 9705160.}
\lref\rVafa{C. Vafa, {\it Evidence for $F$ theory,} Nucl. Phys. B469
(1996) 403, hep-th 9602022.}
\lref\rMe{R. de Mello Koch, work in progress.}

%%%%%%%%%%%%%%%%%%%%%%%%%%%%%%%%%%%%%%%%%%%%%%%%%%%%%%%%%%%%%%%%%%%%%%
%%%%%%%%%%%%%%%%%%          title page       %%%%%%%%%%%%%%%%%%%%%%%%%
%%%%%%%%%%%%%%%%%%%%%%%%%%%%%%%%%%%%%%%%%%%%%%%%%%%%%%%%%%%%%%%%%%%%%%

\Title{CNLS-97-08}
{\vbox {\centerline{Twelve Dimensions and the D2-Brane Tension}
}}

\smallskip
\centerline{Robert de Mello Koch}
\smallskip
\centerline{\it Physics Department and Centre for Nonlinear Studies}
\centerline{\it University of the Witwatersrand}
\centerline{\it Wits 2050, South Africa}\bigskip

\medskip

\noindent

In this letter we study $D$ particle quantum mechanics on a torus in 
the 
limit that one or more cycles of the torus have a zero length.

%%%%%%%%%%%%%%%%%%%%%%%%%%%%%%%%%%%%%%%%%%%%%%%%%%%%%%%%%%%%%%%%%%%%%%

\Date{October, 1997}

%%%%%%%%%%%%%%%%%%%%%%%%%%%%%%%%%%%%%%%%%%%%%%%%%%%%%%%%%%%%%%%%%%%%%%
%%%%%%%%%%%%                text begins                        
%%%%%%%%%%%
%%%%%%%%%%%%%%%%%%%%%%%%%%%%%%%%%%%%%%%%%%%%%%%%%%%%%%%%%%%%%%%%%%%%%%

1) In this letter we study a system of $D$ particles moving 
on a circle. This system arises after compactifying the 
$x^{9}$ co-ordinate of M(atrix) theory \rBFSS\ . We will 
denote the radius of the circle by $R^{9}.$ Consider the 
limit $R^{9}\to 0.$ If we were dealing with field theory 
we would expect to obtain a ten dimensional system, since 
in this limit the momentum excitations in the $x^{9}$ direction 
become heavy and decouple. In string theory the resulting theory 
is still eleven dimensional. This is due to the fact that strings 
winding $R^{9}$ become light in this limit effectively replacing 
the missing momentum modes. Thus the behaviour between string 
theory and field theory in this $R^{9}\to 0$ limit is very 
different. In this letter, we will argue that the $R^{9}\to 0$ 
limit of our system of $D$ particles on a circle yields a 
{\it twelve} dimensional theory. The new twelfth dimension arises 
due to the presence of a two brane in the theory, indicating that 
membrane theory displays yet another behaviour.

This new dimension is in fact conjugate to the two brane tension 
and so appears to be the m(atrix) cousin of the theory discussed 
by Townsend \rTwn\ . In this light, it is natural to expect that 
the limit discussed in this letter will allow the construction of 
manifestly $SL(2,Z)$ invariant m(atrix) string theories. This also 
suggests that there may be a connection between the m(atrix) theory 
we construct in this letter and Vafa's $F$ theory \rVafa\ 
although we have no direct evidence for this.
Moreover, it is clear that configurations with dependence on the 
twelth dimension have a highly nontrivial two brane content. This 
suggests that the theory described in this article may provide a 
number of new insights into non-perturbative brane dynamics. Finally 
we comment on higher dimensional toroidal compactifications of 
m(atrix) theory. In this case it is also natural to expect the 
appearance of new dimensions as the length of certain cycles of the 
torus tend to zero.

2) The Lagrangian which describes the $S^{1}$ compactification is 
given by \rTaylor\ \foot{In this paper we will only deal with the 
bosonic sector of the theory.}

\eqn\SOne
{\eqalign{L&={1\over 2g}\Big[ \sum_{I=1}^{9}\dot{X}_{nI}
\dot{X}^{I}_{-n}-\sum_{J=1}^{8}S_{n}^{J} 
(S_{nJ})^{\dagger}-{1\over 2}\sum_{J,K=1}^{8}
T_{n}^{JK}(T_{nJK})^{\dagger}\Big]\cr
S_{n}^{J}&=\sum_{q}\big[ X_{q}^{9},X_{n-q}^{J}\big]
-2\pi R_{9}nX_{n}^{J},\cr
T_{n}^{JK}&=\sum_{q}\big[ X_{q}^{J},X_{n-q}^{K}\big].}}

The $X^{9}$ coordinate has been compactified on an $S^{1}$ with 
radius $R_{9}.$ Now, lets take the limit $R_{9}\to 0$. After 
taking this limit, the Lagrangian reads

\eqn\ZeroLimit
{L={1\over 2g}\Big[ \sum_{I=1}^{9}
\dot{X}_{nI}\dot{X}^{I}_{-n}
-\sum_{J=1}^{8}S_{n}^{J} (S_{nJ})^{\dagger}
-{1\over 2}\sum_{J,K=1}^{8}
T_{n}^{JK}(T_{nJK})^{\dagger}\Big]}

This Lagrangian contains $9$ space coordinates $X^{I},$ one time 
$t,$ the eleventh dimension which is not manifest and the index 
$n$ which appears as a momentum. The interpretation of the index 
$n$ as a momentum follows from the expectation that the winding 
modes replace the momentum modes as the radius of the theory 
tends to zero. If this interpretation of $n$ as a momentum is 
correct, then we have actually obtained a $12$ dimensional theory! 
Lets start off with a very naive approach and simply take the Fourier 
transform of the above action, treating $n$ as a momentum. The 
transformed Lagrangian has the form

\eqn\TransfrmdAction
{L={1\over 2g}\int dx tr\Big[ 
\sum_{I=1}^{9}\dot{X}_{I}
\dot{X}^{I}
-{1\over 2}\sum_{J,K=1}^{9}
T^{JK}(T_{JK})^{\dagger}\Big]}

Notice that no derivatives with respect to $x$ appear, so that 
classically there does not appear to be any momentum flowing along 
the twelfth dimension. The equations of motion derived from this 
transformed action take the form

\eqn\EqnsNot
{{d^{2}X_{I}\over dt^{2}}-2\big[X^{K},\big[X^{K},X^{I}\big]\big]=0.}

From these equations of motion, it is difficult to imagine how a 
solution with a non-trivial $x$ dependence arises. 

3) There is in fact a simple way in which momentum along the twelfth 
dimension can arise. Consider a family of classical solutions of the 
form

\eqn\ClasSoln
{X^{1}_{n}=\delta_{n,a}xR_{1} {\bf 1},
\quad X^{2}_{n}=(n+a)iR_{2}{d\over dx}{\bf 1} .}

with all other $X_{n}^{I}$ equal to zero.
This solution corresponds, for a specific $n$, to a static $D2$ 
brane in the $1-2$ plane which is easily seen by noting that

\eqn\Commutator
{\big[ X^{J}_{q},X^{K}_{n-q}\big] 
=i\delta_{q,a}nR_{1}R_{2}\delta_{J,1}\delta_{K,2}+
i\delta_{n-q,a}nR_{1}R_{2}\delta_{J,2}\delta_{K,1}.}

The full solution, consisting of all the $X^1_n$ and $X^2_n$ 
corresponds to an infinite number of $D2$ branes stacked in 
the $1-2$ plane. What role does $n$ play? Well, the classical 
energy of the solution (for a given $n$) is 
easily computed to be

\eqn\Energy
{P^{-}={NR_{11}(R_{1}R_{2})^{2}n^{2}\over 4},}

which since $P^{+}=N/R_{11}$ gives rise to a mass

\eqn\massofdt
{m^{2}={(R_{1}R_{2})^{2}N^{2}\over 2}n^{2}.}

Recall that the mass of the brane is given by the area of the brane 
multiplied by the tension of the brane. Secondly, since these branes 
all span the same area (the area of the $1-2$ plane $A=R_{1}R_{2}$), 
we see that the tension of the $n$th solution is proportional to 
$n$. Thus, the twelfth dimension arises as the (conjugate coordinate 
to the) $D2$ brane tension. Since the brane tension enters the 
Lagrangian in exactly the same way as a momentum index, the brane 
tension is conserved in all interactions. This theory is presumably 
the m(atrix) theory version of the manifestly $SL(2,Z)$ invariant 
string action written down by Townsend \rTwn\ . A comment is in 
order. Townsend's action was written for the type $IIB$ string. 
Our system of $D$ particles starts as $IIA$ string theory. 
Interpreting the winding modes as momenta amounts to performing a 
$T$ duality so that our twelve dimensional system is naturally 
interpreted as the type $IIB$ string.

Can this theory be realted to Vafa's $F$ theory. Vafa's theory is a 
$10+2$ dimensional theory. In the above analysis, we have used static 
brane solutions, so that the tension is in fact nothing but the 
energy 
density. In that case, the twelfth dimension which is conjugate to 
$n$ 
(which is an energy density) is naturally interpreted as a time 
dimension. This far from convincing argument suggests that the 
twelfth 
dimension may in fact be timelike.

4) Expanding about the above classical solutions, we are able to 
obtain the
dynamics for fluctuations in the background of these $D2$ branes. The 
action
takes the form

\eqn\Action
{\eqalign{L&={1\over 2g}\int dx tr\Big[ \dot{Y}_{I}\dot{Y}^{I}
+iA\partial_{x}\big[ Y^{2},Y^{1}\big]
+{1\over 2}\big[ Y^{J},Y^{K}\big]\big[ Y_{J},Y_{K}\big]\cr
&-\big[X_{I},Y_{J}\big]^{2}-\big[X_{I},Y_{J}\big]
\big[Y^{I},X^{J}\big]-\big[X_{I},Y^{I}\big]
\big[X_{J},Y^{J}\big]\Big] .}}

If we write $x^{3}=x,$ then the second term in the action can be 
written as

\eqn\ExtraTerm
{\epsilon^{3IJ}\partial_{3}\big[Y^{I},Y^{J}\big].}

This is in the form of a Chern-Simons term in $2+1$ dimensions, which
strongly suggests that this represents an interaction on the world 
volume of the $D2$ brane.\foot{It also suggests that $x^{3}$ is a 
"time" which would be consistent with an $F$ theory interpretation.} 
It would be very interesting to see if the need for this term can be 
traced back to the fact that the two brane couples to the five brane 
\rTwn\ .

5) Finally, consider the case of a compactification of m(atrix) 
theory on a $T^{d}$. Allowing a single radius of the torus to 
tend to zero we obtain exactly the same behaviour as described 
above. However, letting $m$ cycles tend to zero, we obtain an 
$11+m$ dimensional theory. Again, momentum along the extra 
dimensions can be traced back to brane tensions. However, in this
case the classical configurations correspond to branes of increasing 
world volume dimension. It would be interesting to see if the 
Lagrangian \Action\ is manifestly $SL(2,Z)$ invariant (after a 
suitable
compactification) as expected from the similarity between the present
theory and that of \rTwn\ . Presumably then the higher dimensional 
theories obtained when $m$ cycles go to zero would yield theories 
which
are manifestly dual under a larger $U$ duality group. These matters 
will
be addressed in a longer article \rMe\ .

{\it Acknowledgements}$\quad$ It is a real pleasure to thank Jo\~ao 
Rodrigues for helpful discussions.

\listrefs
\vfill\eject
\bye